\documentclass[aps,prl,reprint,groupedaddress]{revtex4-1}

\usepackage[english]{babel}
\usepackage[latin9]{inputenc}
\usepackage{graphicx}

\addto\captionsenglish{}

\usepackage{amsmath, amssymb}

\usepackage{siunitx} 

\usepackage[T1]{fontenc}

\usepackage{hyperref,color}
\definecolor{darkblue}{rgb}{0,.278,.671}
\hypersetup{colorlinks=true, breaklinks=true, linkcolor=darkblue, menucolor=darkblue, urlcolor=darkblue,citecolor=darkblue}

\begin{document}

\title{Field-Dependent Size and Shape of Single Magnetic Skyrmions}

\author{Niklas Romming}
\email[]{nromming@physnet.uni-hamburg.de}
\affiliation{Department of Physics, University of Hamburg, 20355 Hamburg, Germany}
\author{André Kubetzka}
\author{Christian Hanneken}
\author{Kirsten von Bergmann}
\author{Roland Wiesendanger}

\affiliation{Department of Physics, University of Hamburg, 20355 Hamburg, Germany}

\date{\today}

\begin{abstract}
The atomic-scale spin structure of individual isolated skyrmions in an ultrathin film is investigated in real space by spin-polarized scanning tunneling microscopy. Their axial symmetry as well as their unique rotational sense is revealed by using both out-of-plane and in-plane sensitive tips. The size and shape of skyrmions change as a function of magnetic field. An analytical expression for the description of skyrmions is proposed and applied to connect the experimental data to the original theoretical model describing chiral skyrmions. Thereby, the relevant material parameters responsible for skyrmion formation can be obtained.
\end{abstract}


\maketitle

Skyrmions are spatially localized solitonic magnetic whirls with axial symmetry and fixed rotational sense \citep{BOGDANOV1989,Bogdanov1994,Bogdanov1994a,Roessler2011}. They have recently been observed in non-centrosymmetric bulk crystals \citep{Muehlbauer2009,Yu2010,Park2014} as well as in ultrathin transition metal films on heavy-element substrates 
\citep{Heinze2011,Romming2013}, in which a sizable Dzyaloshinskii-Moriya interaction (DMI) \citep{Dzyaloshinsky1958,Moriya1960} induces their formation due to the lack of inversion symmetry. Magnetic skyrmions are in the focus of current research because they offer great potential as information carriers in future robust, high-density, and energy-efficient spintronic devices \citep{Fert2013}: in addition to their protected topology and nano-scale size, they can easily be moved by lateral spin currents \citep{Fert2013,Jonietz2010,Yu2012,Sampaio2013,Nagaosa2013} and written as well as deleted by vertical spin-current injection \citep{Romming2013}. In general, skyrmions couple very efficiently to spin currents and respond sensitively to spin transfer torques. For instance, skyrmion-based racetrack-type memory concepts \citep{Parkin2008} would profit from the fact that skyrmions can be moved with spin-current densities being 5 to 6 orders of magnitude smaller than those needed to move magnetic domain walls \citep{Fert2013}. The fabrication of such skyrmionic devices using ultrathin films and multilayers can stay fully compatible with state-of-the-art technology. While numerous theoretical and simulation studies have concentrated on individual skyrmions and their physical properties, no high-resolution experimental characterization of the internal spin structure of skyrmions has been reported up to now, even though the knowledge about their actual size and shape provides the foundation for predictions about the interactions of skyrmions with spin currents or their manipulation by external fields as envisaged in potential skyrmion-based device concepts. 

In this letter we investigate the magnetic field dependent spin structure of isolated magnetic skyrmions by spin-polarized scanning tunneling microscopy (SP-STM) \citep{Wiesendanger2009}. In contrast to techniques like magnetic force microscopy \citep{Milde2013} and electron holography \citep{Park2014}, the spin structure is measured directly rather than the magnetic field emerging from it. The combination of atomic-scale resolution with spin sensitivity and the ability to measure in external magnetic fields as high as several Tesla makes SP-STM an ideal tool to study the magnetic field-dependent spin structure of single nano-scale skyrmions. The evolution of their size and shape as a function of external magnetic field is connected to micromagnetic theory of skyrmions \citep{Bogdanov1994,Bogdanov1994a} via a proposed analytical description of skyrmions. In this way, the values of the relevant material parameters can be extracted.

\begin{figure} 
\includegraphics{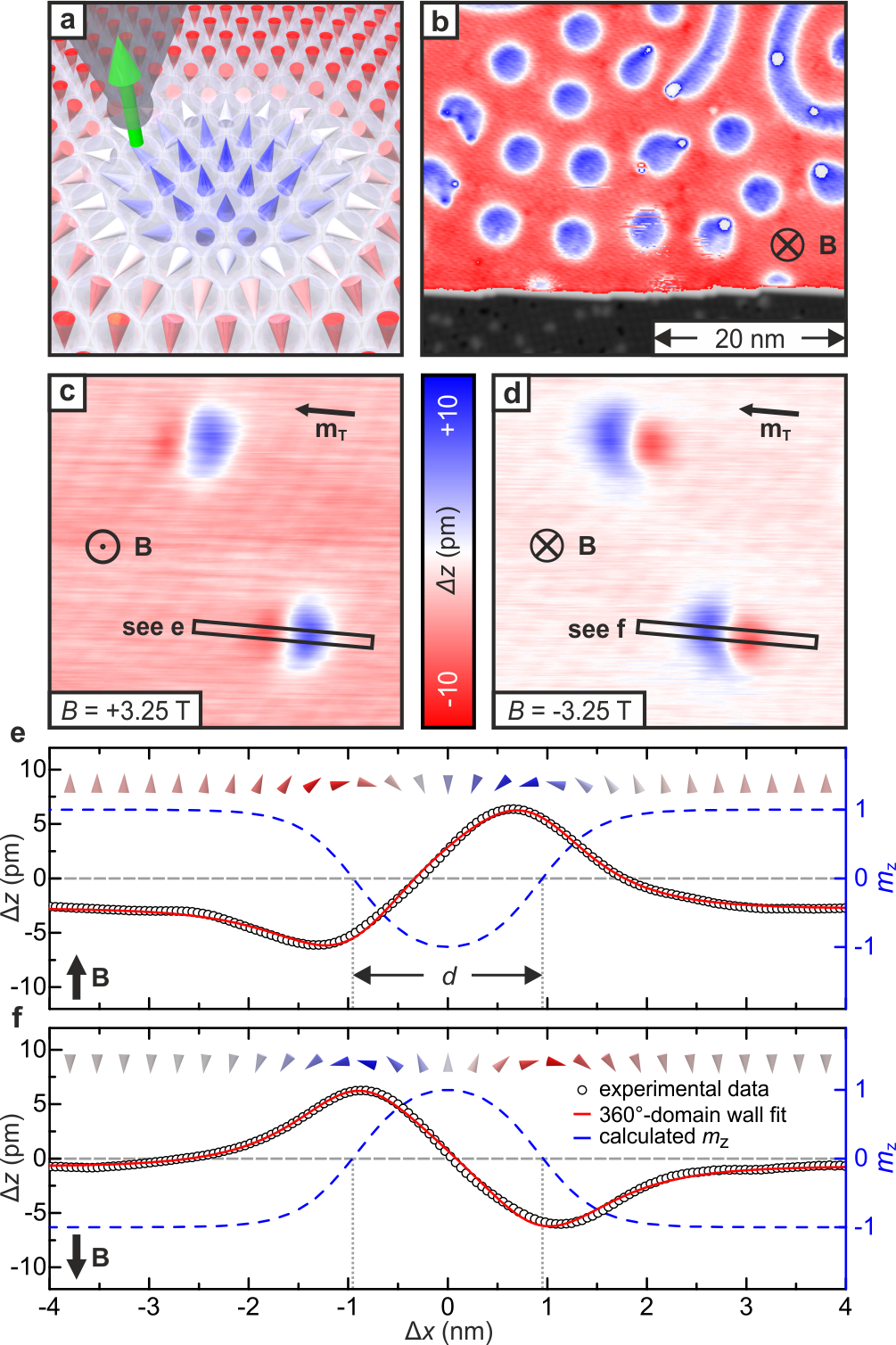} 
\caption{\label{fig1}Spin structure of individual skyrmion in PdFe/Ir(111). (a)~Sketch of the experimental setup of a spin-polarized STM tip probing a magnetic skyrmion. (b)~Topographic constant-current SP-STM image measured with out-of-plane sensitive magnetic tip; each blue circular entity is a skyrmion ($U = \SI[retain-explicit-plus]{+200}{\milli\volt}$, $I = \SI[retain-explicit-plus]{1}{\nano\ampere}$, $T = \SI[retain-explicit-plus]{2.2}{\kelvin}$, $B = \SI[retain-explicit-plus]{-1.5}{\tesla}$). (c)~Magnetic signal (methods \citep{SuppMat}) of two skyrmions measured with an in-plane magnetization of the tip, $\vec{m}_T$, revealing a two-lobe structure ($U = \SI[retain-explicit-plus]{+250}{\milli\volt}$, $I = \SI[retain-explicit-plus]{1}{\nano\ampere}$, $T = \SI[retain-explicit-plus]{4.2}{\kelvin}$). (d)~Same area as in (c) with inverted magnetic field; due to the preserved rotational sense, the contrast is inverted. (e),~(f)~Line profiles across a skyrmion along the rectangles in (c), (d), respectively, and fits with Eq.~(\ref{eq:wall_r}) ((e) $c = \SI{0.90(1)}{\nano\metre}$, $w = \SI{1.18(2)}{\nano\metre}$; (f) $c = \SI{0.91(1)}{\nano\metre}$, $w = \SI{1.17(1)}{\nano\metre}$) and corresponding calculated out-of-plane magnetization $m_z$. The sketches show spins with atomic distance, colorized according to the SP-STM contrast.} 
\end{figure}

Our experimental setup is sketched in Fig.~\ref{fig1}(a): A magnetic probe tip with a well-defined spin orientation at the front atom is used in order to be sensitive to the spin-polarized contribution to the tunnel current, which depends on the projection of the local sample magnetization onto the quantization axis provided by the magnetization direction of the tip. Here, we use an anti-ferromagnetic bulk Cr tip to avoid magnetic interactions of the tip with the sample or an applied field \citep{Kubetzka2002,Schlenhoff2010}.

As sample system we have chosen the bilayer of PdFe on an Ir(111) single crystal substrate, which shows the typical magnetic field induced skyrmion lattice phase \citep{Romming2013}. At the measurement temperature of $T = \SI[retain-explicit-plus]{4.2}{\kelvin}$ the sample exhibits pronounced hysteresis, enabling an investigation of isolated skyrmions in a wide magnetic field range. Fig.~\ref{fig1}(b) shows an overview of a PdFe area, exhibiting several circular skyrmions and two \SI[retain-explicit-plus]{360}{\degree} domain wall sections remaining from the spin spiral phase (top right). Due to the use of an out-of-plane sensitive SP-STM tip, the axisymmetric character of the skyrmions becomes directly evident.

When an in-plane sensitive SP-STM tip is used, the appearance of the skyrmions changes: now two lobes with maximal and minimal spin-polarized current flow are imaged per skyrmion (Fig.~\ref{fig1}(c)) as a result of a positive or negative projection of the local magnetization direction of the skyrmion onto the spin direction of the tip. When tip and sample magnetization directions are orthogonal to each other, the spin-polarized contribution to the tunnel current, and thus the magnetic signal, vanishes: for the image in Fig.~\ref{fig1}(c) this is true close to the center of the Skyrmion and above and below the center. The two skyrmions in the sample area of Fig.~\ref{fig1}(c) appear identical, which is always the case for all skyrmions imaged with a given SP-STM tip. This implies that they exhibit indeed a unique rotational sense \footnote{From our experiments alone we can not identify one of the two rotational senses. Eq.~\eqref{eq:wall_r} and \eqref{eq:Spins} are chosen w.l.o.g.~to be consistent with the skyrmion profiles in Fig.~S1 and a right-handed rotational sense as predicted by Dupé \textit{et al.~}\cite{Dupe2014}.}. According to the symmetry selection rules of the DMI, these interface-induced skyrmions are expected to be cycloidal (sketch in Fig.~\ref{fig1}(a)) \citep{Moriya1960,Bergmann2014}, in agreement with recent density functional theory (DFT) calculations and Monte Carlo simulations for this system \citep{Dupe2014}. When the external magnetic field which induces the skyrmions is applied in the opposite direction, the contrast of the two lobes of the skyrmions is inverted (Fig.~\ref{fig1}(d)) since each spin in the sample is inverted while the spin structure of the anti-ferromagnetic tip remains unchanged, providing an additional proof for the unique rotational sense caused by the DMI. The two skyrmions in Fig.~\ref{fig1}(c) and (d) appear at identical positions due to pinning at atomic defects.

\begin{figure}
	\includegraphics{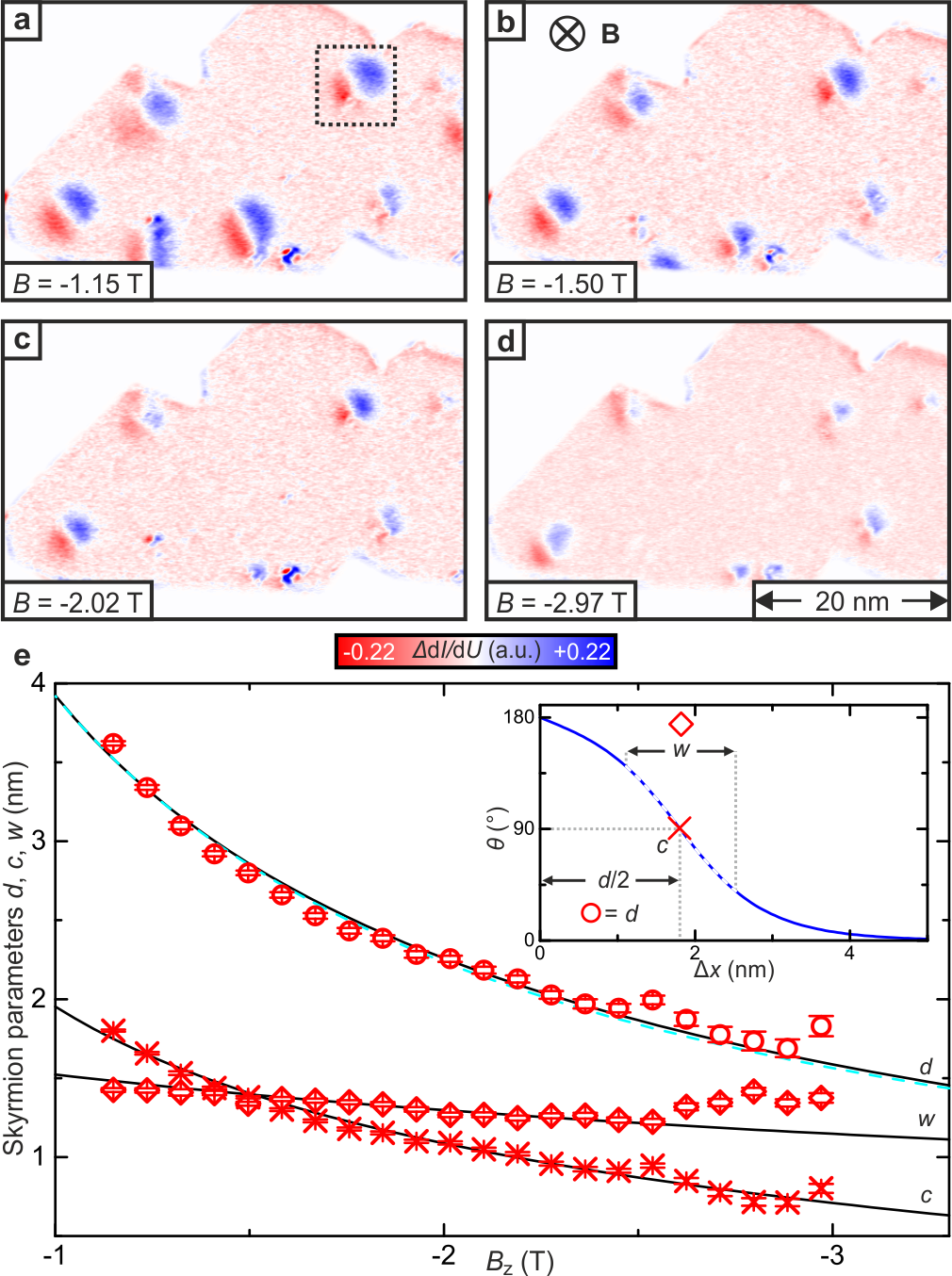}
	\caption{\label{fig2}Evolution of the size and shape of skyrmions in PdFe/Ir(111) as a function of external magnetic field. \mbox{(a)-(d)}~Magnetic signal of SP-STM differential conductance maps (methods \citep{SuppMat}) with in-plane magnetized tip ($U = \SI[retain-explicit-plus]{+20}{\milli\volt}$, $I = \SI[retain-explicit-plus]{3}{\nano\ampere}$, $U_{\text{mod}} = \SI[retain-explicit-plus]{+2.4}{\milli\volt}$, $T = \SI[retain-explicit-plus]{4.2}{\kelvin}$) at the magnetic fields as indicated (Supplementary Movie 1 shows full data set \citep{SuppMat}). (e)~The size and shape of the skyrmion indicated by the box in (a) is evaluated by a fit with Eq.~(\ref{eq:Spins}) as a function of magnetic field. Inset shows geometrical meaning of $c$, $w$ and of $d$, which is numerically calculated; dashed blue line is a fit to $d$ with $1/(B-B_0)$. Solid black lines are obtained theoretically for the fitted set of material parameters $A$, $D$, and $K$ (see text). Error bars correspond to standard deviation of fit parameters.}
\end{figure}

To characterize the size and the shape of a skyrmion we take height profiles across the center (see black rectangles in Fig.~\ref{fig1}(c),(d)). Since there is no exact analytical expression to describe skyrmion profiles, we approximate the cross-section of a skyrmion using a standard \SI[retain-explicit-plus]{360}{\degree} domain wall profile \citep{Braun1994,Kubetzka2003},


\begin{equation}
			\theta(\rho,c,w) = \begin{cases}
			\sum\limits_{+,-}{\left[\arcsin \left( \tanh\frac{-\rho \pm c}{w/2} \right) \right]} + \pi & \text{|} B_z>0\\
		\sum\limits_{+,-}{\left[\arcsin \left( \tanh\frac{-\rho \pm c}{w/2} \right) \right]} & \text{|} B_z<0,
		\end{cases}
	\label{eq:wall_r}
\end{equation}

\noindent where $\theta$ defines the polar angle of the magnetization at position $\rho$, and $c$ and $w$ resemble the position and width of two overlapping \SI[retain-explicit-plus]{180}{\degree} domain walls, respectively. To evaluate the measured data we include the projection of tip and sample magnetization in the fitting procedure (methods \citep{SuppMat}). 

\begin{figure}
	\includegraphics{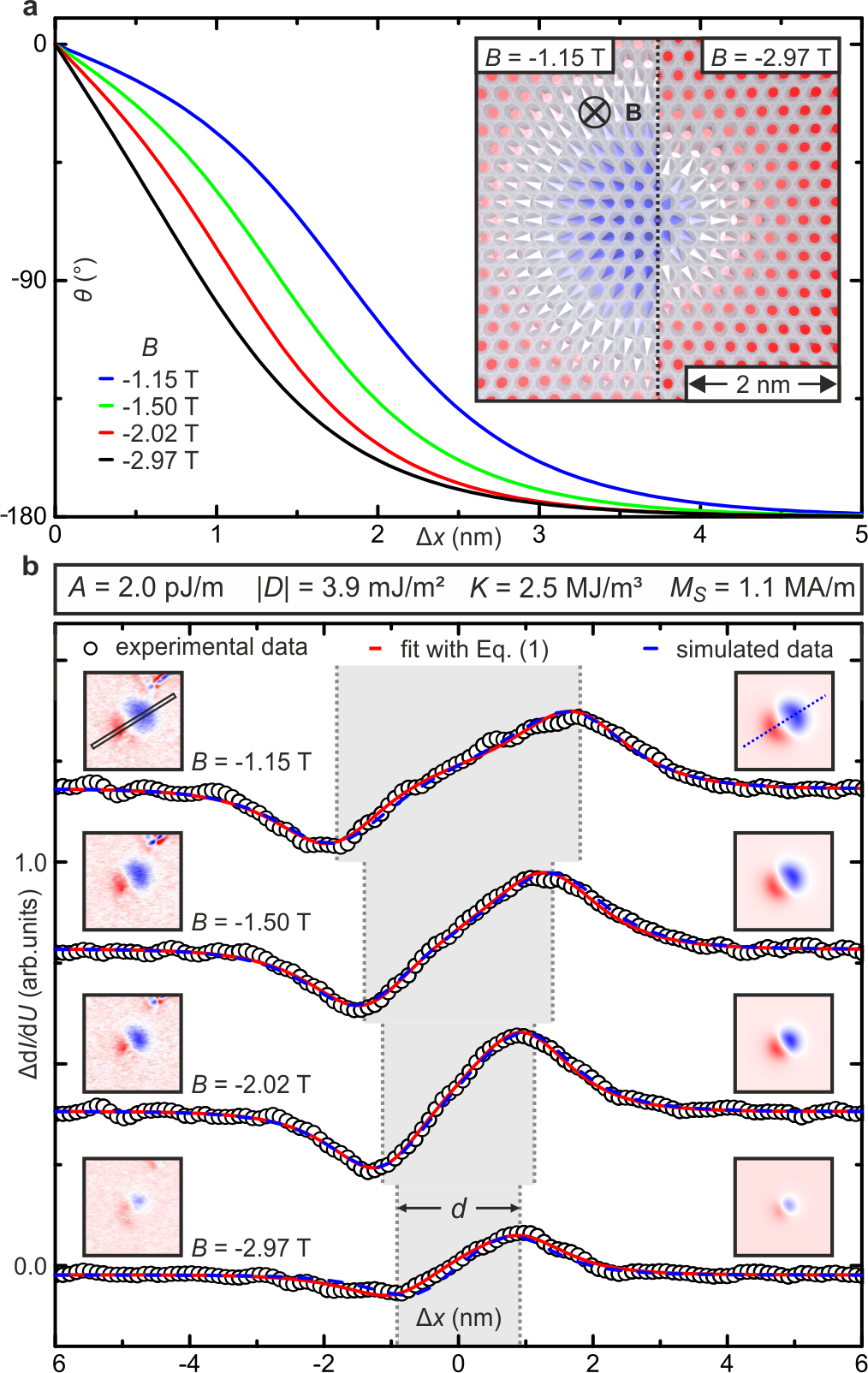}
	\caption{\label{fig3}Validation of material parameters via micromagnetic simulations. (a) Internal spin structure of the skyrmion as described by Eq.~(\ref{eq:wall_r}) for the field values indicated. Inset shows visualization of spins with atomic distance as parametrized by Eq.~(\ref{eq:Spins}); $c$, $w$ given by fits in Fig.~\ref{fig2}(e). (b) Comparison of experimental and simulated height profiles across an individual skyrmion (box in insets) for several magnetic field values, and a fit with Eq.~(\ref{eq:wall_r}). Left and right insets show SP-STM experimental data from \mbox{Fig.~\ref{fig2}(a)-(d)} and micromagnetic simulations based on the derived material parameters, respectively.}
\end{figure}

The agreement between experimental data and fit, see black data points and red fit lines in Fig.~\ref{fig1}(e),(f), justifies the chosen description. Furthermore, a comparison to numerically calculated skyrmion profiles leads to the conclusion that Eq.~(\ref{eq:wall_r}) is an excellent approximation for a wide range of material parameters and field values (Fig.~S1 \citep{SuppMat}). From Eq.~(\ref{eq:wall_r}) it is straightforward to determine the perpendicular magnetization component $m_z(x)$, see blue dashed lines in Fig.~\ref{fig1}(e),(f), and the diameter of the skyrmion $d$, which we define as the diameter of the circle with $m_z = 0$. Exploiting the axial symmetry, the spin structure of an isolated skyrmion in two dimensions is then described by:

\begin{equation}
	\vec{S}(x,y) = \left( \begin{array}{c}
	-\sin(\theta(\rho,c,w)) \cdot x/\rho	\\
		-\sin(\theta(\rho,c,w)) \cdot y/\rho \\
		\cos(\theta(\rho,c,w)) 
	\end{array} \right) ,
\label{eq:Spins}
\end{equation}

\noindent where $\rho = \sqrt{x^2+y^2}$ is the radial distance from the center of the skyrmion located at the origin. Note that within this model the spin structure of the skyrmion is fully determined by only two parameters, $c$ and $w$.

The impact of external magnetic fields onto the size and shape of a skyrmion becomes evident in field-dependent SP-STM experiments: Fig.~\ref{fig2}\mbox{(a)-(d)} show the identical sample area imaged at different external magnetic field strengths. The color scale resembles the magnetic contribution to maps of differential tunnel conductance (d$I$/d$U$), and the use of an in-plane sensitive tip leads again to the two-lobe appearance of the skyrmions. The decrease of the skyrmion size with increasing field can be directly seen in the displayed image sequence (Supplementary
Movie 1 shows full data set \citep{SuppMat}). For a quantitative analysis we fit a single isolated skyrmion (black box in Fig.~\ref{fig2}(a)) with our two-dimensional skyrmion model and obtain the characteristic parameters $c$, $w$, and $d$. 

The evolution of these parameters with external field is shown in Fig.~\ref{fig2}(e). The diameter of the skyrmion roughly scales with $1/(B-B_0)$, see dashed blue line, in agreement with numerical calculations \citep{Bogdanov1994a,Wilson2014}. While, in the investigated field range, the diameter changes by more than a factor of two, the effect on the width of the transition region is only about $\SI[retain-explicit-plus]{25}{\percent}$. Consequently, as can be seen in Fig.~\ref{fig3}(a), the skyrmion shape changes qualitatively with magnetic field, leading to a significant decrease in the number of spins with a component opposite to the magnetic field. This results from a subtle balance of all involved energies, where the Zeeman energy leads to a compressing force and the DMI stabilizes skyrmions against collapse to the ferromagnetic state.

To assess these underlying interactions for the biatomic PdFe layer in the framework of micromagnetic continuum theory, we establish a connection to the standard energy functional in cylindrical coordinates \citep{BOGDANOV1989,Bogdanov1994a,Bogdanov1994,Wilson2014,Bogdanov1999,Rohart2013}:

\begin{align*}
	E  = 2 \pi t \int_0^\infty  \Bigg[ &  A  \left( \left( \frac{\text{d} \theta}{\text{d} \rho} \right) ^2 + \frac{\sin^2 \theta}{\rho^2} \right)  \\
	 & + D  \left( \frac{\text{d} \theta}{\text{d} \rho} + \frac{\sin \theta  \cos \theta}{\rho} \right) \tag{3}\\
	 & - K  \cos^2 \theta - B_z  M_S  \cos \theta  \Bigg]  \rho \text{d} \rho 
	\label{eq:Energy}
\end{align*}

\noindent where exchange stiffness $A$, DMI constant $D$, uniaxial effective anisotropy constant $K$, and saturation magnetization $M_S$ are the material dependent parameters, $B_z$ is the external out-of-plane magnetic field and $t$ is the film thickness. From DFT calculations \citep{Dupe2014}, we estimate an $M_S \approx \SI[retain-explicit-plus]{1.1}{\mega\ampere\per\metre}$  (methods \citep{SuppMat}). The magnetization profile $\theta(\rho)$ is given by our experimentally verified skyrmion model, Eq.~(\ref{eq:wall_r}). Now for each set of $A$, $D$, $K$, $B$ the energy functional can be minimized with respect to $c$ and $w$. These theoretical curves $c(B)$ and $w(B)$ are fitted to the experimentally obtained values for $c(B)$ and $w(B)$ via an error weighted least square fit with $A$, $D$, $K$ as fitting parameters (Fig.~\ref{fig2}(e)). The solid lines are the calculated values of $d$, $c$, $w$ for $A=\SI[retain-explicit-plus]{2.0(4)}{\pico\joule\per\metre}$,  $D=\SI[retain-explicit-plus]{3.9(2)}{\milli\joule\per\metre\squared}$ and $K=\SI[retain-explicit-plus]{2.5(2)}{\mega\joule\per\metre\cubed}$ \footnote{For a detailed discussion on the error bars, see Supplemental Material \cite{SuppMat}.} as a function of magnetic field. These parameters are in the range expected for thin-film systems \citep{Dupe2014,Simon2014}, and the agreement of $c$ and $w$ obtained from theory with those from a fit to the experimental data is evident.

To demonstrate that these derived material parameters can be used to accurately reproduce the experimental data, we perform micromagnetic simulations \citep{Rohart2013,OOMMF,SuppMat}. Fig.~\ref{fig3}(b) shows height profiles across an isolated skyrmion at four different magnetic field values. The SP-STM data (black circles), the skyrmion fit with Eq.~(\ref{eq:wall_r}) (red line) and the height profile across the skyrmion in a micromagnetic simulation (blue dashed) nicely coincide, and the real-space agreement between experimental data and simulation is demonstrated in the insets to Fig.~\ref{fig3}(b). Additionally, the accurate description of the field-dependent magnetism of the PdFe bilayer by the derived material parameters extends to lower magnetic fields (Fig.~S2 \citep{SuppMat}).

The presented SP-STM study provides access to the actual spin structure of an isolated skyrmion, enabling a direct comparison to micromagnetic theory. Since the field-dependent evolution of size and shape of single skyrmions is governed by a balance of the magnetic interactions, a detailed experimental characterization can yield the relevant material parameters such as the DMI, which is responsible for the stability of these particle-like states. This procedure is applicable to all DMI-stabilized magnetic objects; however, for thick layers or small DMI the demagnetizing field should be taken into account explicitly to obtain realistic material parameters. We regard the precise knowledge of the spin structure of skyrmions as a prerequisite for further explorations of their possible application in spintronic devices.

\begin{acknowledgments}
We thank A. N. Bogdanov, S. Rohart and J. Hagemeister for discussions. We acknowledge financial support from the Deutsche Forschungsgemeinschaft (DFG) via grants SFB668 and GrK 1286.
\end{acknowledgments}


%

\end{document}